\documentclass[trackchanges, twocolumn]{aastex7}
\usepackage{amsmath}
\usepackage{gensymb}
\usepackage[utf8]{inputenc}
\usepackage[T1]{fontenc}

\usepackage{hyperref}



\DeclareRobustCommand{\ion}[2]{%
  \relax\ifmmode
    \ifx\testbx\f@series
      {\mathbf{#1\,\mathsc{#2}}}%
    \else
      {\mathrm{#1\,\textsc{#2}}}%
    \fi
  \else
    \textup{#1\,{\mdseries\textsc{#2}}}%
  \fi}

\begin{document}

\title{Discovery of Isolated, Quenched Candidate Backsplash Dwarf Galaxies near M101}

\author[orcid=0009-0008-7952-5702,sname='North America']{Julian Shapiro}
\affiliation{The Dalton School, 108 E 89th St, New York, NY 10128, USA}
\email{c26js@dalton.org}

\vspace{0.3cm}

\accepted{to ApJ}

\begin{abstract}

I report the discovery of three faint, semi-resolved quiescent candidate dwarf galaxies, which are potential backsplash galaxies associated with the nearby spiral M101 ($D \sim 6.7 \,\mathrm{Mpc}$). The galaxies are cataloged as Shapiro DG-I/MAGE1412+5650, Shapiro DG-II, and Shapiro DG-III. Sha DG-I is concurrently discovered in ID-MAGE. The backsplash candidates lie in a magnitude range $M_V\sim-7.3$ to $-8.1$, and half-light radii $r_h\sim110-140\,\mathrm{pc}$. Hydrodynamical simulations suggest a population of backsplash galaxies that have been environmentally stripped by interactions with a host and ejected from the system, though they have not yet been definitively observed in the local universe. Initial distance estimates are determined using surface brightness fluctuations in Sérsic model-subtracted CFHTLS and HSC archival images. The galaxies are not detected in GALEX, or H$\alpha$ for Sha DG-I/MAGE1412+5650, providing upper limits on their SFR history. The provisional distances and quiescence of the dwarf candidates are consistent with a potential backsplash nature, though HST/JWST follow-up are necessary for stronger constraints, and to probabilistically distinguish from cosmic web stripping. Sha DG-II is alternatively a possible satellite of NGC 5585. Abundance models of satellites based on halo masses, which can be extended to model backsplash galaxies, will enable important consistency tests of $\Lambda\mathrm{CDM}$ with upcoming surveys, e.g., LSST. This is briefly explored in the context of the M101 group.

\end{abstract}

\keywords{\uat{Galaxies}{573} --- \uat{Dwarf galaxies}{416} --- \uat{Galaxy evolution}{594} --- \uat{Quenched galaxies}{2016} --- \uat{Galaxy quenching}{2040} --- \uat{Galaxy structure}{622}}

\section{Introduction} 

Understanding the quiescence of low-luminosity dwarf galaxies, both satellites and in isolation, is crucial for addressing the $\Lambda$ Cold Dark Matter ($\Lambda$CDM) model and its challenges at a small scale (e.g. \citealt{Klypin_1999,Moore_1999,10.1111/j.1745-3933.2011.01074.x,10.1111/j.1365-2966.2012.20695.x,annurev:/content/journals/10.1146/annurev-astro-091916-055313,annurev:/content/journals/10.1146/annurev-astro-091918-104453}). Thus, low-mass dwarfs have been a theoretical focus through increasingly high-resolution cosmological simulations (e.g. \citealt{10.1093/mnras/stw145,Wetzel_2016,10.1093/mnras/stx1710,10.1093/mnras/stz3054,Applebaum_2021}). Progress has been made in reconciling discrepancies between simulations and observed populations within the LG; however, it is necessary to look to other systems in the Local Volume (LV; $D\leq10 \,\mathrm{Mpc}$, \citealt{Karachentsev_2013}) with extensive optical surveys (e.g. \citealt{Martin_2013,Geha_2017,10.1093/mnras/sty1772,Drlica-Wagner_2020,Mao_2021,Carlsten_2022,10.1093/mnrasl/slac063,Mao_2024}) revealing many new faint dwarf galaxy candidates.

In addition to satellite dwarfs, isolated low-mass dwarfs provide a test of galaxy evolution far from the influence of hosts, and they have only recently been explored in optical surveys (e.g. \citealt{10.1111/j.1365-2966.2012.21581.x,McQuinn_2015,Sand_2022,Jones_2023,Jones_2024,Li_2024,Sand_2024}). Their level of quiescence can be compared to characteristics expected in hydrodynamical simulations (e.g. \citealt{Dickey_2021}). LG observations (e.g. \citealt{Weisz_2014}) support the well-studied morphology-density relation (e.g. \citealt{1984ApJ...281...95P}), showing a robust correlation between the distance to the host and quiescence. Dwarfs within the virial radii of their hosts are generally quenched and gas-poor due to environmental gas stripping, while dwarfs outside of the virial radii are typically gas-rich and detected in \ion{H}{i} (e.g. \citealt{Putman_2021}). Spherical collapse models by \citet{Gunn:1972sv, 1984ApJ...281....1F, 1985ApJS...58...39B} defined the splashback mechanism, where cold shells reach an apocenter after first infall. These backsplash galaxies have orbits that pass them within $0.5R_{200}$, where they are environmentally stripped, and are then ejected from their hosts' dark matter halos to beyond $R_{200}$. This quenching mechanism has been proposed as a means of explaining isolated, quiescent dwarf galaxies \citep[e.g.][]{10.1111/j.1365-2966.2012.21793.x}, and is a focus of high-resolution simulations \citep[e.g.][]{More_2015,10.1093/mnras/sty2913,Benavides_2021}.


Numerous previous searches have identified low luminosity dwarf galaxy candidates near M101 \citep{Merritt_2014,Karachentsev2015,Bennet_2017,Carlsten_2022,refId0b,refId02}, a bulgeless spiral with a stellar mass similar to the MW ($M_\star \approx 5.3 \times 10^{10} M_\odot$; \citealt{van_Dokkum_2014}), though with a likely lower dark matter halo mass and a stellar halo smaller than the MW's. Recent estimates of its distance from Cepheids find $D=6.7\pm0.1\,\mathrm{Mpc}$ \citep{Riess_2024}. Its satellite system has been considered complete down to $M_\star \approx 5 \times 10^{5} M_\odot$ within the virial radius of M101 ($\sim 250 \,\mathrm{ kpc}$; \citealt{Merritt_2014}), corresponding to a projected radius $\sim2.1\degree$ where the nine that have been confirmed as true satellites and not background low surface brightness (LSB) galaxies all reside.

Here, I present the discovery of three low-luminosity, quiescent candidate dwarf galaxies in the outskirts of the M101 group. Considering simulation predictions of how backsplash galaxies are distributed and the lack of studies outside of M101's virial radius, a search was carried out in the $\sim250\mathrm{-}350\,\mathrm{kpc}$ region for faint, gas-poor backsplash dwarfs, resulting in the discovery of the candidates. In Section 2, I present archival and follow-up observations. Derived structural properties and analysis are provided in Section 3. In Sections 4--6, I discuss their candidacy as backsplash dwarfs and outline future follow-up.

\begin{figure*}[t]
\centering
    \includegraphics[width=1\linewidth]{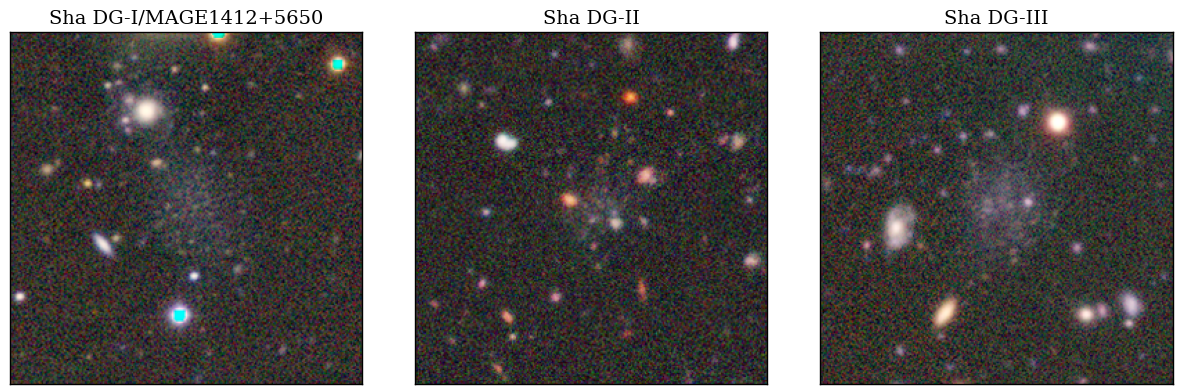}
    \caption{$z/i,r,g$ cutout images of Shapiro DG-I/MAGE1412+5650 (left), Shapiro DG-II (center), and Shapiro DG-III (right) from Subaru Hyper Suprime-Cam and the Canada-France-Hawaii Telescope Legacy Survey. N is up and E is left. The FOV is 54 $\times$ 54 arcsec.}
    \label{fig:enter-label}
\end{figure*}

\section{Observations and Archival Data}

\subsection{Hyper Suprime-Cam Survey}

Shapiro DG-I/MAGE1412+5650 and DG-II were discovered in archival $z$ band data from the Hyper Suprime-Cam (HSC; \citealt{10.1093/pasj/psx063}) during a search for M101 backsplash candidates. Sha DG-I was discovered concurrently by \citet{Hunter_2025} in the ID-MAGE survey for potential satellites of LMC/SMC-mass hosts, in this case NGC 5585, and is cataloged as MAGE1412+5650. Shapiro DG-III was discovered in the Legacy Archive of the Hyper Suprime-Cam (HSCLA; \citealt{10.1093/pasj/psab034}).

Subaru observations provide sufficient angular resolution and depth to reveal semi-resolved stellar populations within the new candidate dwarfs. From visual inspection of the calibrated HSC products, the candidates show dwarf spheroidal (dSph) morphologies typical of quiescent dwarfs observed around MW analog hosts within their virial radii, and lack star-forming regions.

\subsection{CFHTLS}

The archival Canada-France-Hawaii Telescope Legacy Survey (CFHTLS; \citealt{Gwyn_2012}), obtained from the Canada Astronomy Data Centre (CADC), provides $u,g,r,i,z$ coverage of the backsplash candidates. The images are of lower resolution than HSC. However, the CFHTLS data are of higher $\mathrm{S/N}$ in some regions and provide a more complete photometric wavelength range. Image depths and exposure lengths of the wide layer observations used in this work are shown in Table \ref{tab:cfhtls_table}. 

A $\sim 9\,\mathrm{deg}^2$ area of the CFHTLS survey field, approximately centered on M101, was used by \citet{Merritt_2014,Bennet_2017} to identify LSB dwarf candidates within the $\sim2.1\degree$ virial radius of M101 and does not overlap with this work.


\subsection{Galaxy Evolution Explorer}

The UV sky survey from the Galaxy Evolution Explorer (GALEX; \citealt{Martin_2005}), retrieved using the Barbara A. Mikulski Archive for Space Telescopes (MAST), is used to search for coincident emissions from the backsplash candidates. NUV images of 1536 s (MIS), 3363.85 s (GII), and 25898.55 s (DIS), and FUV images of 96 s (AIS), 96 s (AIS), and 4024.15 s (DIS) are used for Sha DG-I/MAGE1412+5650, Sha DG-II, and Sha DG-III respectively. No UV emission associated with the galaxies was detected in any of the GALEX observations.



\subsection{Apertif}

The archival first data release of the APERture Tile In Focus (Apertif; \citealt{refId0apertif}) imaging survey \citep{refId0apertif2} from the Westerbork Synthesis Radio Telescope is used to search for potential associated \ion{H}{i}. The survey has previously been used with dwarf galaxies by \citet{refId0apertif3}, who determined \ion{H}{i} morphologies from a sample of typical stellar mass dwarf galaxies within $M_\star$ $\approx$ $10^{8}-5 \times 10^{9}M_\odot$. The \ion{H}{i} spectral cubes of the regions containing the galaxies were accessed from the ASTRON Virtual Observatory Data Center. The field from beam 35 was used, providing coverage of Sha DG-I/MAGE1412+5650 and Sha DG-II with a frequency range of $1414.5 \,\mathrm{MHz}\mathrm{-}1429.3 \,\mathrm{MHz}$. No resolved coincident sources were found within the Apertif survey data.

\subsection{Two-meter Twin Telescope}

I collected H$\alpha$ observations of Sha DG-I/MAGE1412+5650 for this work using the 80 cm TTT2 Ritchey--Chrétien telescope of the Two-Meter Twin Telescope facility located at the Teide Observatory. The telescope, operating at f/4.4, is equipped with a QHY411M camera containing a 151 MP sensor. With a pixel size of 3.76 $\mu \mathrm{m}$, the pixel scale of the system is $0.22\,\mathrm{arcsec\, pixel^{-1}}$. The H$\alpha$ and Sloan $r$ filters were used to perform continuum subtraction. Observations were taken over a period of four nights, with $102 \times 300\,\mathrm{ s}$ collected in H$\alpha$ and $34 \times 120\,\mathrm{ s}$ collected in Sloan $r$. The images were automatically flat-field calibrated and astrometrically corrected. No detection of Sha DG-I/MAGE1412+5650 was made in the $\mathrm{H}\alpha$ or lower integration Sloan $r$ images.

\section{Measurements and Properties}


\subsection{Structural Properties}

\begin{deluxetable*}{lccc}\label{structural}
\tabletypesize{\scriptsize}
\tablewidth{\textwidth} 
\tablecaption{Properties of the Dwarf Candidates\label{table2}}
\tablehead{
  \colhead{Parameter}   & \colhead{Sha DG-I} & \colhead{Sha DG-II} & \colhead{Sha DG-III}
}
\startdata
$\alpha_0$ (J2000)           &14:12:11.9&14:16:32.3&14:13:13.5\\
$\delta_0$ (J2000)          &+56:50:39&+56:22:31&+52:23:09\\
$r_\mathrm{h}$ (arcsec)       &6.8 $\pm$ 0.8 &3.5 $\pm$ 0.1&3.2 $\pm$ 0.2\\
$r_\mathrm{h}$ (pc)           &140 $\pm$ 80&110 $\pm$ 60&130 $\pm$ 20\\
$n_\text{Sérsic}$           &0.5 $\pm$ 0.1&0.5 $\pm$ 0.2&0.4 $\pm$ 0.1\\
$\epsilon$           &0.3 $\pm$ 0.1&0.3 $\pm$ 0.1&0.2 $\pm$ 0.1\\
$\mathrm{PA}$&$\sim$21$\degree$&$\sim$122$\degree$&$\sim$177$\degree$\\
\hline
$m-M$ (mag)   &$28.1^{+1.0}_{-0.7}$&$29.0^{+1.0}_{-1.0}$&$29.5^{+0.3}_{-0.3}$\\
$D_{\odot}$ (Mpc)     &$4.2^{+2.4}_{-1.2}$&$6.2^{+3.7}_{-2.3}$&$8.0^{+1.0}_{-0.9}$\\
$D_\mathrm{M101,\,proj.}$ (kpc)&329&324&289\\
$M_V$& $-7.3^{+0.7}_{-1.0}$&$-7.5^{+1.0}_{-1.0}$&$-8.1^{+0.3}_{-0.3}$\\
$\mathrm{SFR}_\mathrm{NUV}$ ($M_\odot \,\mathrm{yr}^{-1}$)&$\lesssim4.7 \times 10^{-6}$&$\lesssim4.5 \times 10^{-7}$&$\lesssim9.2 \times 10^{-8}$\\
$\mathrm{SFR}_\mathrm{FUV}$ ($M_\odot \,\mathrm{yr}^{-1}$)&$\lesssim6.3 \times 10^{-4}$&$\lesssim4.0 \times 10^{-5}$&$\lesssim1.7 \times 10^{-6}$\\
$M_\text{\ion{H}{i}}\,(M_\odot)$&$\lesssim1.4\times 10^{5}$&$\lesssim3.0 \times 10^{5}$&---\\
\enddata
\tablecomments{The half-light radius $r_\mathrm{h}$ [pc], $M_{V}$, SFR, and M\textsubscript{\ion{H}{i}} are distance-dependent and use SBF estimates, carrying their uncertainties. As distances are not precisely constrained and require follow-up, the derived properties should be treated as provisional.}
\end{deluxetable*}

\begin{figure}
\centering
    \includegraphics[width=1\linewidth]{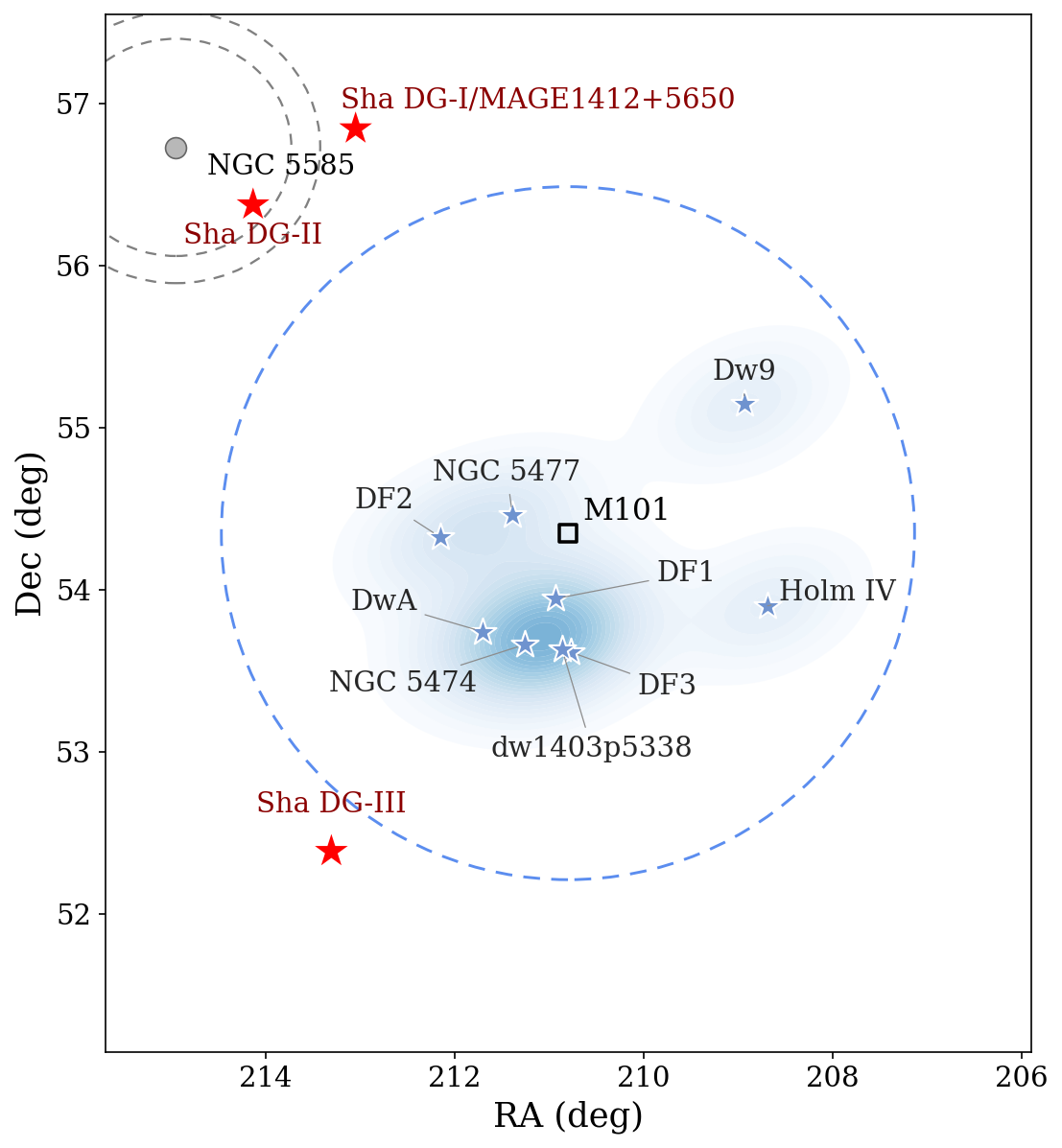}
    \caption{Map centered on M101 (center box) of the projected positions of associated galaxies in a $\sim400 \times400\,\mathrm{ kpc}$ field. The confirmed M101 dwarf satellites are shown as blue stars. The kernel density estimate of the satellites is shown as the blue to white contour. The candidate dwarf galaxies discovered in this work are highlighted as the red stars. The virial radius of M101, $\sim250 \,\mathrm{ kpc}$, is displayed as the blue dotted circle surrounding M101. Unrelated background galaxies from \citet{Bennet_2017,Bennet_2019} are not shown. The virial radius of NGC 5585 (dotted grey circle) is estimated based on its mass \citep{Hunter_2025}.}
    \label{fig:enter-label}
\end{figure}

\begin{figure*}
\centering
    \includegraphics[width=1\linewidth]{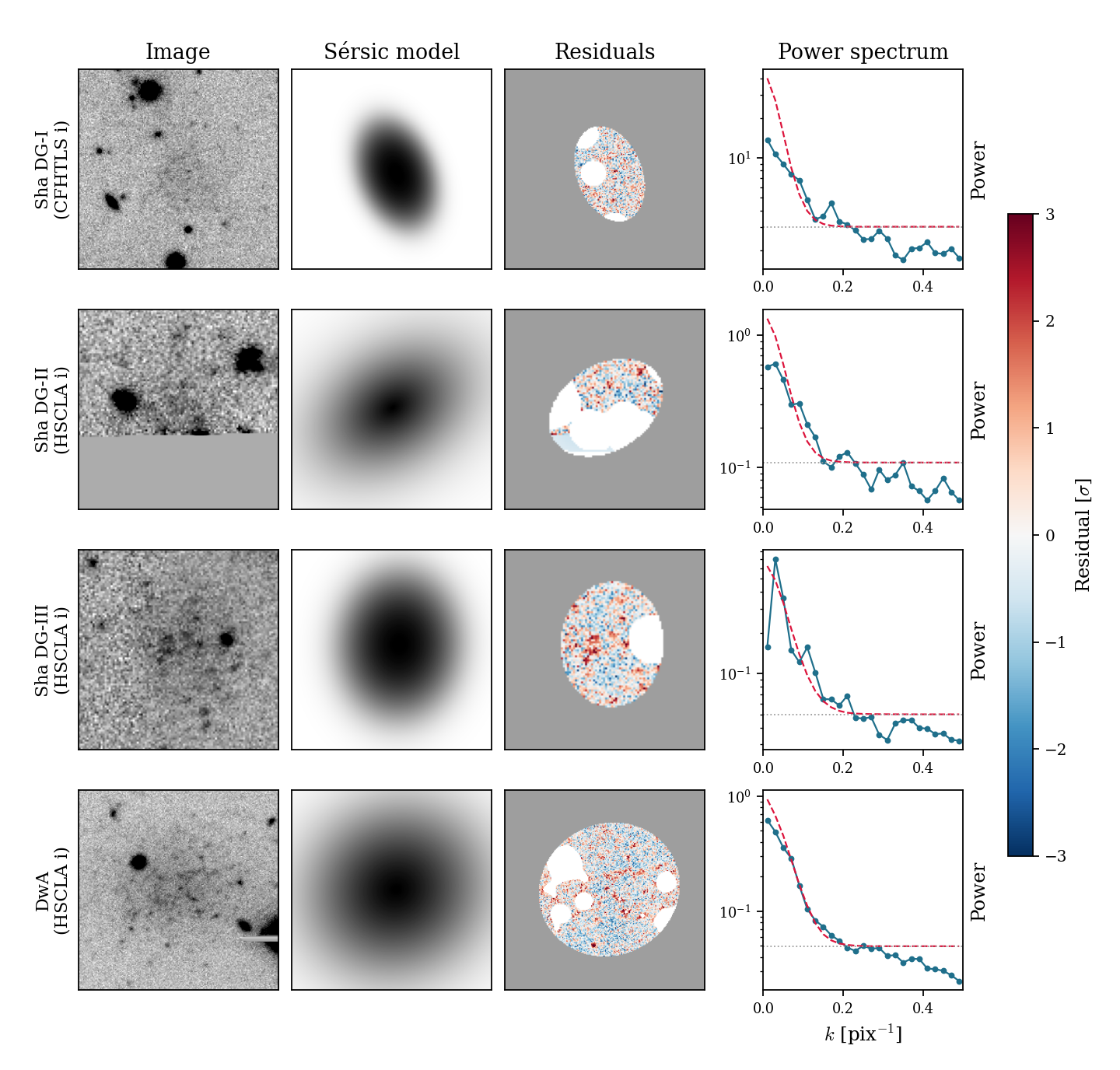}
    \caption{Surface brightness fluctuation models of the three dwarf candidates and DwA (used as a consistency check). The first column displays the CFHTLS or HSCLA i-band cutouts of the galaxies. The second column shows the best-fit Sérsic models. The model-subtracted residuals used for SBF are shown in the third column, with a colorbar in units $\sigma$. The last column displays the power spectra with the model $P(k)$ as a dotted red line. The observed spectrum is the dotted blue line, and the noise level is shown in grey. The HSCLA Sérsic model of Sha DG-II is elongated as it overlaps with the coverage edge; this does not affect its SBF distance, which is in agreement with CFHTLS coverage. The CFHTLS model is used instead for $n_\text{Sérsic}$ and $\epsilon$.}
    \label{fig:sbf}
\end{figure*}

The candidates are located beyond the $\sim2.1\degree$ virial radius of M101 at projected separations of $\sim2.81\degree$ for Sha DG-I/MAGE1412+5650, $\sim2.77\degree$ for Sha DG-II, and $\sim2.47\degree$ for Sha DG-III. The data are corrected for Galactic extinction using \citet{Schlegel_1998} maps with \citet{Schlafly_2011} coefficients. The $E(B-V)$ reddening values are 0.0096, 0.0121, and 0.0157 for the candidates. The $V$ band magnitudes of the candidates are calculated using the conversions from \citet{refId0c}. Best-fit Sérsic models are created using Astropy after masking contaminants, resulting in values of n\textsubscript{Sérsic} = $0.5 \pm 0.1$, $0.5 \pm 0.2$, and $0.4 \pm 0.1$ for the three candidates, respectively. The ellipticity ($\epsilon$) is determined with image moments in NumPy, with a Gaussian filter applied to reduce noise, deriving $\epsilon = 0.3 \pm 0.1$, $0.3 \pm 0.1$, and $0.2 \pm 0.1$, in line with known M101 quenched dwarfs DwA ($\epsilon = 0.33 \pm 0.06$) and Dw9 ($\epsilon \leq 0.33$) \citep{Bennet_2019}. The half-light radii of the candidates are estimated as $r_\mathrm{h}= 140 \pm 80$ pc, $110 \pm 60$ pc, and $130 \pm 20$ pc from the models.

\subsection{Distance}
The $z$ and $i$ band luminosity functions of the detected stellar populations in the dwarf candidates lie in a range $\sim24\mathrm{-}27$. A precise TRGB is not attempted due to the limited number of point sources and ambiguity regarding whether stars are in the AGB or RGB. However, the detection is sufficient to exclude the possibility of the galaxies belonging to the LG, where both AGB and RGB stars are $i\,\mathrm{mag}\approx20-23$ (e.g. \citealt{Lee_2021}). An upper limit $\sim16\,\mathrm{Mpc}$ can be placed based on Subaru's point source depth $i\approx26\mathrm{-}27$.

Surface brightness fluctuations (SBF) are a well-established method of estimating extragalactic distances, using luminosity fluctuations from unresolved stars contributing to the pixel fluxes (e.g. \citealt{Tonry1988, Tonry_2001}). SBF has been demonstrated to be effective with dwarf galaxies in archival survey data, reproducing TRGB to $\sim15\%$ accuracy around LV hosts \citep{Carlsten_2020}. \citet{Carlsten_2019} directly uses SBF and CFHTLS to distinguish likely satellites from unassociated background galaxies in the M101 group. Following the methodology of \citet{Cantiello_2018, Carlsten_2020, Li_2026}, residual images are produced by subtracting Sérsic models from the CFHTLS and HSCLA images. Contaminant point sources are masked out, and a model $P(k)$ is produced as $P_0E(k)+P_1$, where $P_0$ is the SBF fluctuation amplitude, $E(k)$ is the expectation power spectrum, and $P_1$ is the image noise level. $g\mathrm{-}i$ is assumed $\approx0.70\pm0.15$, which is a dominant source of systematic error in $D$. The cutouts, models, residuals, and power spectra of the candidates are displayed in Figure \ref{fig:sbf}. As a consistency check, the methodology is also used on DwA to determine $D\approx6.6\pm0.9\,\mathrm{Mpc}$, reproducing TRGB by \citet{Bennet_2019}. The resulting distance estimates for the three new candidate dwarfs are provided in Table \ref{structural}. Values should be treated as provisional, considering the survey depth limitations. 

\subsection{Quiescence}

The GALEX UV flux is a strong tracer of a relatively young stellar population ($\lesssim100\,\mathrm{Myr}$) and has been used in works surrounding low-luminosity, semi-resolved dwarfs to assess their quiescence \citep[e.g.][]{Sand_2022}. Galaxies with strong NUV and FUV continuum have experienced abundant star formation in the past $100\,\mathrm{Myr}$ \citep{Leroy_2019}. No flux in either band is detected in the candidate backsplash dwarf galaxies. Using the methodology described by \citet{10.1093/mnras/stad3521}, and the UV SFR convention of \citet{annurev:/content/journals/10.1146/annurev-astro-081811-125610,Leroy_2019}, $3\sigma$ upper limits are derived as $\mathrm{SFR}_\mathrm{NUV}\lesssim 4.7\times10^{-6}\,,\,4.5\times10^{-7},\,9.2\times10^{-8}M_\odot \,\mathrm{yr}^{-1}$ and $\mathrm{SFR}_\mathrm{FUV}\lesssim 6.3\times10^{-4},\,4.0\times10^{-5},\,1.7\times10^{-6}\,M_\odot \,\mathrm{yr}^{-1}$ for the three candidate dwarfs, respectively. Similarly, non-detections of the dwarf candidates in Apertif allow limits to be placed on the \ion{H}{i} content of Sha DG-I/MAGE1412+5650 and DG-II, which lie within the survey coverage. Using a channel width of $12.15 \,\mathrm{kHz}$ and a line width of $20 \,\mathrm{km\, s^{-1}}$, the $3\sigma$ upper limits are $M_{\text{\ion{H}{i}}} \lesssim 1.4 \times 10^{5}M_{\odot}$ and $M_{\text{\ion{H}{i}}} \lesssim 3.0 \times 10^{5}M_{\odot}$, respectively.

The limits placed on the GALEX UV SFR and \ion{H}{i} content are consistent with quenching in the candidates. Further, the galaxies' low Sérsic indices, dwarf spheroidal (dSph) morphologies, and low ellipticities are in line with that of known isolated quiescent dwarfs in the Local Volume (e.g. Tucana B, \citealt{Sand_2022}; Hedgehog, \citealt{Li_2024}; Sculptor A, \citealt{Sand_2024}).

\begin{figure}[ht!]
\centering
    \includegraphics[width=1\linewidth]{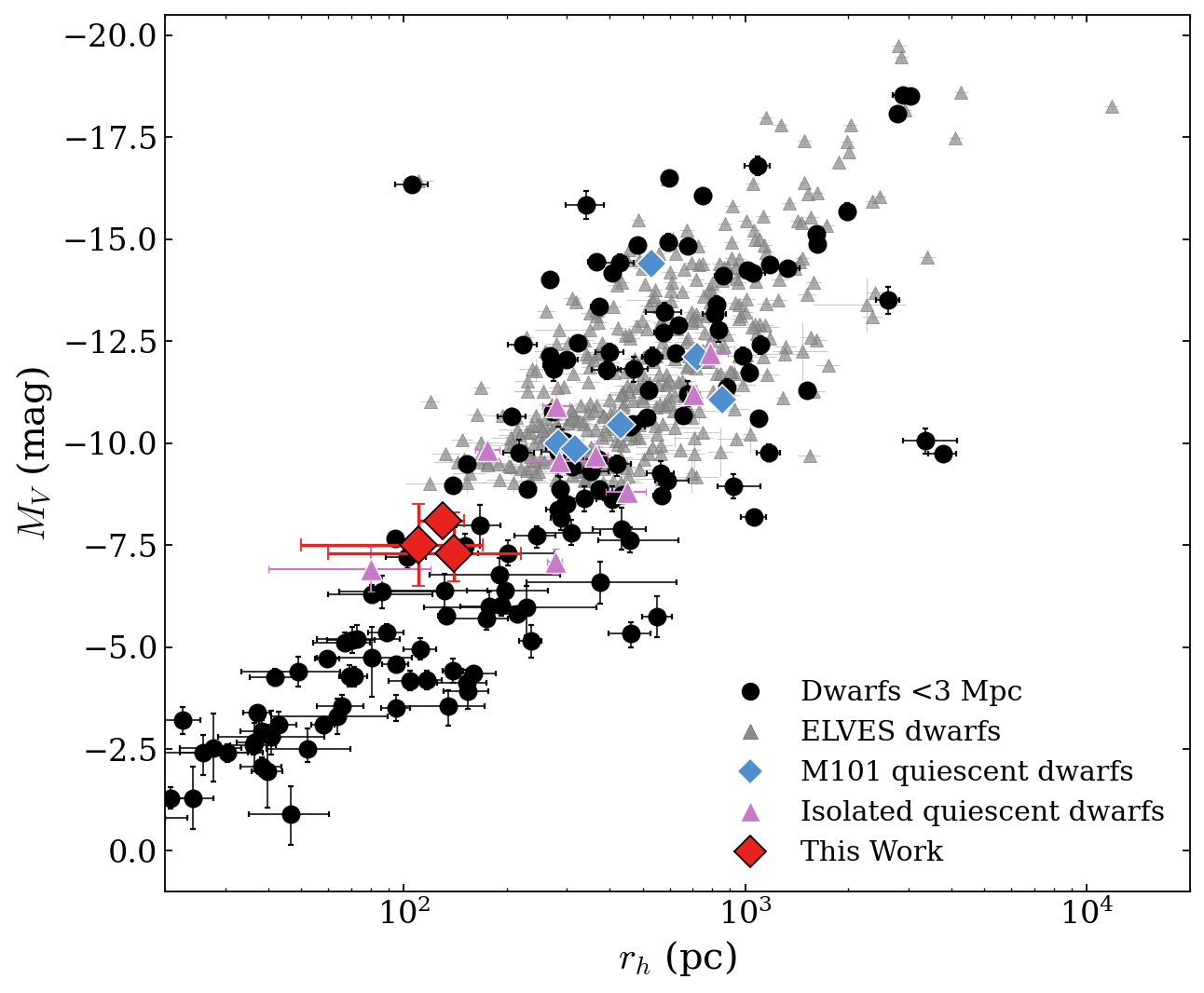}
    \caption{The $V$ band magnitudes ($M_V$) and half-light radii ($r_\mathrm{h}$) of dwarf galaxies within the Local Volume. The candidates from this work are shown alongside known M101 quiescent dwarfs \citep{Bennet_2019}, isolated and quiescent dwarfs (Cetus, \citealt{10.1111/j.1365-2966.2005.09806.x}; Tucana, \citealt{1996A&A...315...40S}; KKR25, \citealt{10.1111/j.1365-2966.2012.21581.x}; And XVIII, \citealt{McConnachie_2008}; Eri II, \citealt{Crnojević_2016b}; COSMOS-dw1, \citealt{Polzin_2021}; Tucana B, \citealt{Sand_2022}; Blobby, \citealt{10.1093/mnras/stad3521}; Hedgehog, \citealt{Li_2024}), dwarfs within $3\,\mathrm{Mpc}$ \citep{McConnachie_2012}, and dwarfs from the Exploration of Local VolumE Satellities (ELVES; \citealt{Carlsten_2022}) survey. }
    \label{fig:enter-label}
\end{figure}

\section{Environment}

\subsection{Backsplash Scenario}

Observations from high-resolution optical surveys such as HSC-SSP and the DESI Legacy Imaging Surveys have revealed a handful of quenched dwarfs outside of the virial radii of hosts. Both simulations (e.g. \citealt{10.1093/mnras/sty774}) and surveys predict these objects to be rare, $\lesssim0.06\%$ of field dwarfs \citep{Geha_2012}. Two viable quenching mechanisms exist for this population: ejected backsplash dwarfs that previously experienced ram-pressure stripping, and cosmic web stripping, where isolated galaxies cross overdense regions, creating a challenge in determining the nature of individual candidates. The SBF distance estimates are consistent with the $\sim1.5\,\mathrm{Mpc}$ expected range for backsplash galaxies, but the possibility of cosmic web stripping is not excluded due to the candidates' distance uncertainty range. Sha DG-II also lies in the projected virial radius of NGC 5585, and may instead be a faint satellite. Though Sha DG-I is an ID-MAGE candidate satellite of NGC 5585, it is outside of its virial radius. Probabilistically distinguishing the nature of each candidate will require distance measurements from HST or JWST, discussed in Section \ref{follow-up}.

\section{Towards Backsplash $\Lambda$CDM Tests}\label{sec5}

As future observations will construct a larger population of local backsplash candidate galaxies, comparisons with empirical galaxy formation models will be critical to examine consistency with $\Lambda$CDM. While a dedicated analysis is outside the scope of this paper, initial tests are examined in the context of the M101 group, and can help inform backsplash abundance models and their relationship to host properties and satellites. 3.6 and 4.5 $\mu \mathrm{m}$ model flux densities from \citet{Eskew_2012} are used by \citet{10.1093/mnras/stae2809} to obtain a stellar mass of $M_\star = 5.5 \pm0.52\times 10^{10} M_\odot$ for M101. The total dark matter halo mass is often estimated from abundance matching (AM), with a monotonic double power-law stellar-to-halo mass relation (SHMR): \citep{10.1111/j.1365-2966.2009.15268.x,Moster_2010,Girelli2020}:
\begin{equation}
\frac{M_\star}{M_\mathrm{h}}=2A(z)
\left[
\left(\frac{M_\mathrm{h}}{M_A(z)}\right)
^{-\beta(z)}+
\left(\frac{M_\mathrm{h}}{M_A(z)}\right)^{\gamma(z)}\right]^{-1}
\end{equation}
resulting in $M_\mathrm{h}\approx1.3\times 10^{12}M_\odot$ for M101 with \citet{Girelli2020} parameters. This is inconsistent with kinematic measurements, which find $M_\mathrm{h}\approx4\mathrm{-}8.5\times 10^{11}M_\odot$ from radial velocities and luminosity distributions of known satellites \citep{Tikhonov2015,Karachentsev2019,10.1093/mnras/stab2621}.

\citet{Posti2021} finds that for masses $M_\star\lesssim5\times10^{10}M_{\odot}$, AM models consistently over-predict $M_\mathrm{h}$, preventing accurate satellite population estimates. Instead, I use an isothermal sphere model (e.g. \citealt{Kormendy_2016}) with $\sigma=88 \,\mathrm{ km \,s}^{-1}$ \citep{1987A&A...179...23A} to estimate $M_\mathrm{h}\approx6.74\times10^{11}M_\odot$, consistent with kinematic measurements. The estimated subhalo count within $M_\mathrm{peak}=0.5\mathrm{-}3\times10^9M_\odot$, where dwarfs are expected to form \citep{10.1093/mnras/stv1691}, is derived for the MW and M101 using the equation \citep{10.1111/j.1365-2966.2010.17601.x}: 
\begin{equation}
N(>\mu\equiv m_\mathrm{sub}/M_{200})=\left(\frac{\mu}{\widetilde\mu_1}\right)^a \exp
\left[
-\left(\frac{\mu}{\mu_\mathrm{cut}}
\right)^b
\right]
\end{equation}
to find a factor of $\sim2.3$ difference in expected satellites between the MW and M101, consistent with the $\sim2.5$ factor discrepancy between average MW-like hosts and M101 described by \citet{Bennet_2019}.

Examining the consistency of observed satellite abundances with physical parameters on a wide scale is a critical small-scale test of $\Lambda$CDM and can be expanded to similarly test future backsplash populations. In addition to halo mass, the tidal index parameter $\Theta_5$ \citep{Karachentsev_2013} evaluates environmental density:
\begin{equation}
\Theta_5=\log\left(\sum^5_{n=1}M_n/D^3_\mathrm{in}
\right)+C,
\end{equation}
and can be used alongside $M_h$ in relation to satellite abundances:
\begin{equation}
M_{h,w}\equiv M_h10^{k\Theta_5}
\end{equation}
The resulting correlation of $\sim4.6\sigma$ among a LV sample \citep{10.1093/mnras/stu329,Lelli_2016} is stronger than either parameter individually, hinting that it may be a useful consistency test for observations of satellite and backsplash populations (e.g. with Rubin LSST; \citealt{jurić2015lsstdatamanagement,bosch2018overviewlsstimageprocessing}).

\begin{figure}
\centering
    \includegraphics[width=1\linewidth]{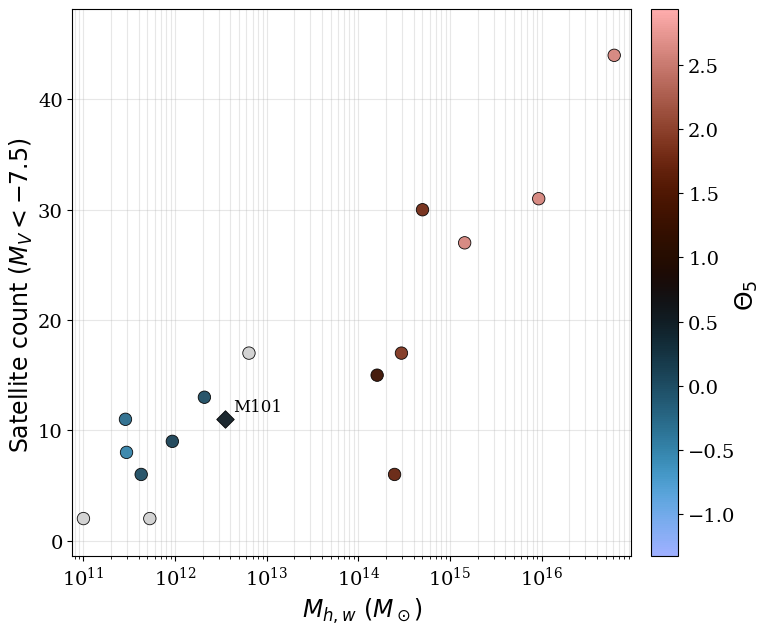}
    \caption{Number of satellites $M_V<-7.5\,\mathrm{mag}$ around nearby MW-like galaxies as a function of density contrast-weighted mass $M^{k\Theta_5}_\mathrm{h}M_\odot$, where $M_\mathrm{h}$ is the kinematically derived halo mass, and $\Theta_5$ is the tidal index from the five closest neighbors. The best-fit slope is 10.19 and the best-fit weight is $k=0.77$. The colors display $\Theta_5$ on its own for the sample. The Pearson correlation is $r=0.889$ ($\sim4.6\sigma$). The included galaxies are M31, NGC 1023, NGC 253, NGC 2683, NGC 2903, NGC 3031, NGC 3115, NGC 4258, NGC 4594, NGC 4631, NGC 4736, NGC 5055, NGC 5236, M101, NGC 628, and NGC 891. NGC 2903 may be an outlier, with a relatively isolated environment but a high tidal index. M101 fits well in the trend.}
    \label{fig:tidalindex}
\end{figure}

\section{Future Follow-up}\label{follow-up}

Recent HST studies of quiescent, isolated LG dwarfs have found that none of the six observed galaxies are likely to be backsplash within $6\,\mathrm{Gyr}$, with only Cetus remaining a plausible candidate \citep{bennet2025orbitsisolateddwarfslocal}, despite expectations of a substantial LG backsplash population \citep{10.1093/mnras/sty2913}. Further follow-up is needed beyond the LG to examine whether observed backsplash populations agree with simulations. Imaging of the new candidates in this work from HST or JWST would provide direct TRGB distance measurements to probabilistically assess the possibility of a backsplash origin. Ground-based spectra from Keck may enable studies of their radial velocities. If confirmed to be backsplash, they would be a unique and ideal testing ground to explore the masses, stellar population ages, and tidal histories of this classification of galaxy.

\section{Conclusions}

This paper presents the discovery of three isolated, quiescent dwarf galaxy candidates, cataloged as Shapiro DG-I/MAGE1412+5650, Shapiro DG-II, and Shapiro DG-III. Sha DG-I was concurrently discovered by \citet{Hunter_2025}. Archival $z$ and $i$ band observations from the Canada France Hawaii Telescope and Hyper Suprime-Cam are used to create Sérsic models of the candidates. The luminosity function of the semi-resolved stellar populations excludes the possibility of very nearby Local Group members, and provides an upper limit of $\sim16\,\mathrm{Mpc}$ from the Subaru point source depth limit. The surface brightness fluctuations of the Sérsic-subtracted images are used to obtain initial distance estimates $\sim4\mathrm{-}8\,\mathrm{Mpc}$, consistent with potential backsplash members, though deeper observations are required to better constrain their nature. The models are used to find the $V$ band magnitudes of $\sim-8.1$ to $-7.3$, and the half-light radii $r_h\sim110-140\,\mathrm{pc}$, similar to known isolated quenched dwarfs. The values for the Sérsic indices fall in a range $\sim0.4\mathrm{-}0.5$, and the ellipticities are in a range $\sim0.2\mathrm{-}0.3$. Nondetections in GALEX, and additionally a H$\alpha$ nondetection in Sha DG-I, show a lack of recent star formation and support a quiescent nature. The candidates' quiescence and SBF-estimated proximity to M101 hint that they may be backsplash, having previously been environmentally stripped. Cosmic web stripping, or a faint satellite of NGC 5585 in the case of Sha DG-II, remains a possibility. Deeper follow-up observations with HST or JWST would enable far stronger distance constraints, providing the opportunity to probabilistically assess their candidacy as backsplash galaxies. Evolving high-resolution optical surveys, such as the Rubin LSST, are expected to find a large population of backsplash dwarf candidates around LV hosts. The expected abundances of satellites in M101, based on dark matter halo mass, are briefly explored in this work. Accurate abundance models will be necessary to extend to backsplash galaxies in LSST, enabling critical small-scale comparisons to $\Lambda\mathrm{CDM}$ expectations.
 
\begin{acknowledgments}

I thank Dr. Patrick Ogle for helpful comments on the manuscript.I thank the anonymous reviewer for important feedback to improve the final paper.

This article includes observations made in the Two-meter Twin Telescope (TTT) in the Teide Observatory of the IAC, that Light Bridges operates in the Island of Tenerife, Canary Islands (Spain). The Observing Time Rights (DTO) used for this research were provided by the project GALAXDIF25.

The Hyper Suprime-Cam (HSC) collaboration includes the astronomical communities of Japan and Taiwan, and Princeton University. The HSC instrumentation and software were developed by the National Astronomical Observatory of Japan (NAOJ), the Kavli Institute for the Physics and Mathematics of the Universe (Kavli IPMU), the University of Tokyo, the High Energy Accelerator Research Organization (KEK), the Academia Sinica Institute for Astronomy and Astrophysics in Taiwan (ASIAA), and Princeton University. Funding was contributed by the FIRST program from the Japanese Cabinet Office, the Ministry of Education, Culture, Sports, Science and Technology (MEXT), the Japan Society for the Promotion of Science (JSPS), Japan Science and Technology Agency (JST), the Toray Science Foundation, NAOJ, Kavli IPMU, KEK, ASIAA, and Princeton University. This paper makes use of software developed for Vera C. Rubin Observatory. We thank the Rubin Observatory for making their code available as free software at https://pipelines.lsst.io/. This paper is based on data collected at the Subaru Telescope and retrieved from the HSC data archive system, which is operated by the Subaru Telescope and Astronomy Data Center (ADC) at NAOJ. Data analysis was in part carried out with the cooperation of Center for Computational Astrophysics (CfCA), NAOJ. We are honored and grateful for the opportunity of observing the Universe from Maunakea, which has the cultural, historical and natural significance in Hawaii.

The Pan-STARRS1 Surveys (PS1) and the PS1 public science archive have been made possible through contributions by the Institute for Astronomy, the University of Hawaii, the Pan-STARRS Project Office, the Max Planck Society and its participating institutes, the Max Planck Institute for Astronomy, Heidelberg, and the Max Planck Institute for Extraterrestrial Physics, Garching, The Johns Hopkins University, Durham University, the University of Edinburgh, the Queen’s University Belfast, the Harvard-Smithsonian Center for Astrophysics, the Las Cumbres Observatory Global Telescope Network Incorporated, the National Central University of Taiwan, the Space Telescope Science Institute, the National Aeronautics and Space Administration under grant No. NNX08AR22G issued through the Planetary Science Division of the NASA Science Mission Directorate, the National Science Foundation grant No. AST-1238877, the University of Maryland, Eotvos Lorand University (ELTE), the Los Alamos National Laboratory, and the Gordon and Betty Moore Foundation.


This paper is based in part on data from the Hyper Suprime-Cam Legacy Archive (HSCLA), which is operated by the Subaru Telescope. The original data in HSCLA was collected at the Subaru Telescope and retrieved from the HSC data archive system, which is operated by the Subaru Telescope and Astronomy Data Center at National Astronomical Observatory of Japan. The Subaru Telescope is honored and grateful for the opportunity of observing the Universe from Maunakea, which has the cultural, historical and natural significance in Hawaii. 



Based in part on observations obtained with MegaPrime/MegaCam, a joint project of CFHT and CEA/IRFU, at the Canada-France-Hawaii Telescope (CFHT) which is operated by the National Research Council (NRC) of Canada, the Institut National des Science de l'Univers of the Centre National de la Recherche Scientifique (CNRS) of France, and the University of Hawaii. This work is based in part on data products produced at Terapix available at the Canadian Astronomy Data Centre as part of the Canada-France-Hawaii Telescope Legacy Survey, a collaborative project of NRC and CNRS.

This research used the facilities of the Canadian Astronomy Data Centre operated by the National Research Council of Canada with the support of the Canadian Space Agency.

This research is based in part on observations made with the Galaxy Evolution Explorer (GALEX), obtained from the MAST data archive at the Space Telescope Science Institute, which is operated by the Association of Universities for Research in Astronomy, Inc., under NASA contract NAS 5–26555.

This work makes use of data from the Apertif system installed at the Westerbork Synthesis Radio Telescope owned by ASTRON. ASTRON, the Netherlands Institute for Radio Astronomy, is an institute of the Dutch Research Council (“De Nederlandse Organisatie voor Wetenschappelijk Onderzoek, NWO).

This research has made use of the NASA/IPAC Extragalactic Database (NED), which is operated by the Jet Propulsion Laboratory, California Institute of Technology, under contract with the National Aeronautics and Space Administration.

\end{acknowledgments}


%
\facilities{Subaru, Canada-France-Hawaii Telescope (CFHT), GALEX, Westerbork Synthesis Radio Telescope (WSRT), Two-meter Twin Telescope (TTT)}

\software{astropy \citep{2013A&A...558A..33A, 2018AJ....156..123A, 2022ApJ...935..167A},  
          photutils \citep{2024zndo..10967176B},
          numpy \citep{harris2020array},
          matplotlib \citep{Hunter:2007},
          plotly \citep{plotly_2024_14503524}
          }


\appendix

\section{Luminosity Functions}

\begin{figure}[ht!]
    \centering
    \includegraphics[width=1\linewidth]{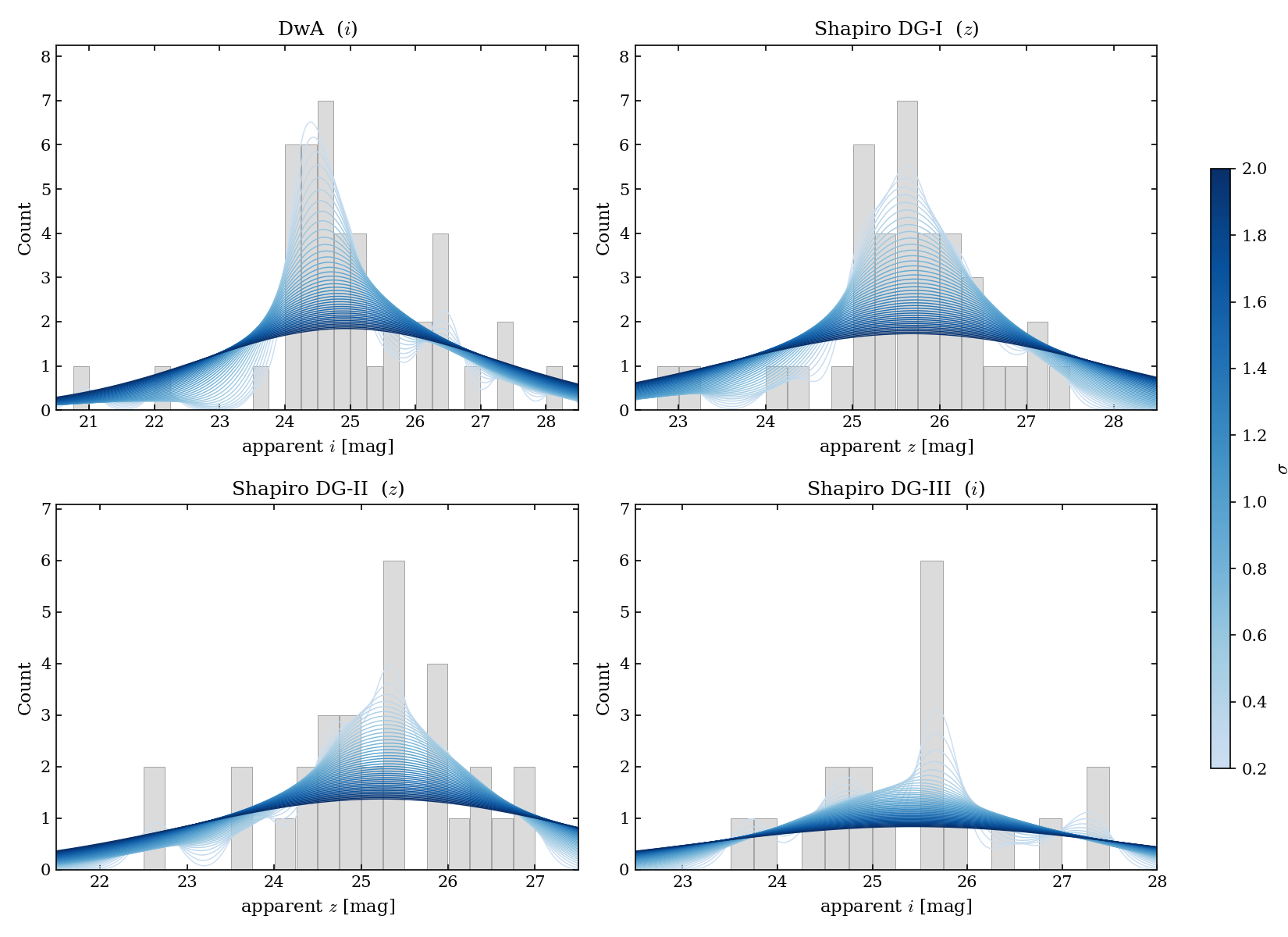}
    \caption{The $z$ and $i$ band luminosity functions of the semi-resolved stellar populations detected in Subaru images. The blue curves show Gaussian smoothed functions in $\sigma$.}
    \label{fig:lf}
\end{figure}

The detected semi-resolved stellar populations in the candidate dwarf galaxies are used to create the $z$ and $i$ band luminosity functions shown in Figure \ref{fig:lf}. The luminosity functions are not used to directly measure distance, but are consistent with populations of known M101 satellites such as DwA.

\section{Unresolved Candidates}

\begin{figure}[ht]
\centering
    \includegraphics[width=0.5\linewidth]{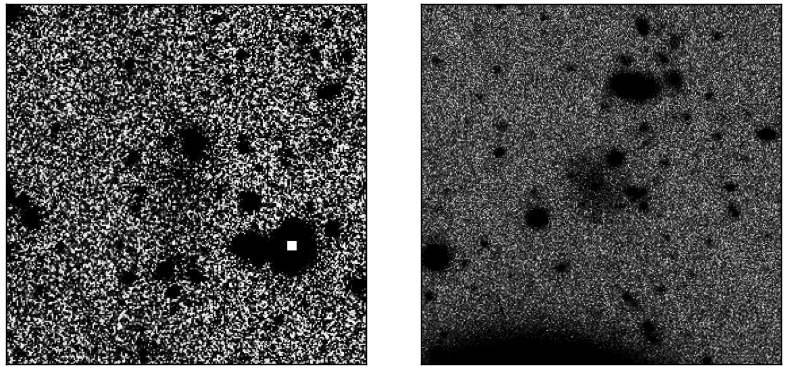}
    \caption{HSC $z$ and $i$ band images of additional candidates not selected due to the lack of a clear resolved stellar population.}
    \label{fig:additional}
\end{figure}

From the search, two additional faint, quiescent dwarf galaxies, dw1426+5718 and dw1419+5443, were identified in the HSC data and are shown in Figure \ref{fig:additional}. These galaxies are within the projected distance range from M101 to be backsplash; however, resolved populations are not identified to the limits of the HSC image depths, and require follow-up to determine their membership.

\section{CFHTLS Pointings}

The exposure lengths and 50\% completeness depths $m_{\mathrm{lim}}$ for each of the CFHTLS pointings for the candidates in $u^*,\,g',\,r',\,i',\,z'$ are provided in Table \ref{tab:cfhtls_table}.

\begin{deluxetable*}{lccccccccccc}
\tabletypesize{\scriptsize}
\tablewidth{\textwidth} 
\tablecaption{CFHTLS Archival Observations\label{tab:cfhtls}}

\tablehead{
  \colhead{Pointing} &
  \colhead{Candidate} &
  \multicolumn{2}{c}{$u^*$} &
  \multicolumn{2}{c}{$g'$} &
  \multicolumn{2}{c}{$r'$} &
  \multicolumn{2}{c}{$i'$} &
  \multicolumn{2}{c}{$z'$} \\
  \cline{3-4} \cline{5-6} \cline{7-8} \cline{9-10} \cline{11-12}
  \colhead{} &
  \colhead{} &
  \colhead{$m_\mathrm{lim}$} & \colhead{$t_{\rm exp}$ (s)} &
  \colhead{$m_\mathrm{lim}$} & \colhead{$t_{\rm exp}$ (s)} &
  \colhead{$m_\mathrm{lim}$} & \colhead{$t_{\rm exp}$ (s)} &
  \colhead{$m_\mathrm{lim}$} & \colhead{$t_{\rm exp}$ (s)} &
  \colhead{$m_\mathrm{lim}$} & \colhead{$t_{\rm exp}$ (s)}
}

\startdata
W3-1+3 & Sha DG-I/MAGE1412+5650 &
26.1 & 3001.1 &
26.3 & 2501.0 &
25.7 & 2000.8 &
---  & ---        &
24.6 & 3601.0 \\
W3-1+2 & Sha DG-I/MAGE1412+5650 &
25.7 & 3001.0 &
26.7 & 2501.0 &
25.8 & 2001.1 &
25.6 & 4306.7 &
24.5 & 3601.2 \\
W3+0+2 & Sha DG-II &
25.6 & 3001.0 &
26.4 & 2501.0 &
25.9 & 2001.1 &
25.6 & 4306.5 &
24.4 & 3601.3 \\
W3-1-2 & Sha DG-III &
25.3 & 3001.0 &
26.3 & 2501.0 &
26.0 & 2501.1 &
25.4 & 4306.5 &
24.6 & 3601.2 \\
\enddata

\tablecomments{50\% completeness depths $m_\mathrm{lim}$ \citep{Gwyn_2012} and exposure times $t_\text{exp} \text{ (s)}$ of the CFHTLS pointings containing the candidates.
}
\label{tab:cfhtls_table}
\end{deluxetable*}

\nocite{*}
\bibliography{sample7}{}
\bibliographystyle{aasjournal}



\end{document}